# The Architects of Narrative Evolution: Actor Interventions Across the SAGES Framework in Information Campaigns

Lynnette Hui Xian Ng
Carnegie Mellon University, USA
lynnetteng@cmu.edu

Yukai Zeng
Ministry of Defence, Singapore
zeng_yukai@mindef.gov.sg

Muthiah Ponmani
Ministry of Defence, Singapore
muthiah_so_ponmani@mindef.gov.sg

**Abstract**

*Narratives in digital spaces are not merely organic phenomena. They are strategically shaped by a range of actors to influence public perception, behavior, and sociopolitical outcomes. This paper offers an actor-oriented expansion of the SAGES Framework, a five-stage model that traces the evolution of narratives from digital inception to real-world impact: Seeding, Amplification, Galvanization, Expansion, and Stickiness. This framework maps how adversarial and constructive actors intervene at each stage to accelerate, redirect, or counter narrative trajectories. Through comparative case studies of the 2021 Myanmar military coup and the 2022 Russia–Ukraine war, we show how narrative manipulation campaigns unfold and how targeted interventions can mitigate their effects. The SAGES framework contributes a practical lens for analyzing influence operations and developing countermeasures in an era of contested information ecosystems.*

## 1. Introduction

In today's contested information environments, narratives are not merely viral byproducts of organic discourse — they are strategically shaped by actors who seed, accelerate, or suppress narratives to influence public perception, behavior, and ultimately, societal norms. These actors range from state-linked propaganda units and troll farms to fact-checkers, civic organizations, and platform moderators. Actors function as architects of narrative evolution, orchestrating influence campaigns that blur the boundaries between online manipulation and offline consequences.

This analysis contributes to ongoing research on adversarial digital influence and the erosion of institutional trust in online ecosystems. The viral spread of false information can create dangerous echo-chambers of conspiracy theories and hoaxes, change people's stances against the truth and can spark physical violence (Ng et al., 2022; Phillips et al., 2022). Therefore, much attention in the social-cybersecurity realm has been devoted towards the profiling such information campaigns (Carley, 2020). Prior work typically examined digital information campaigns situated in the digital realm. Cyber security implementations have been profiled with the MITRE ATTA&CK and D3FEND Frameworks (Menard et al., 2025), the disinformation lifecycle with the C5Integration Framework (Kruijver et al., 2025), and the analysis of digital propaganda through the BEND framework (Marigliano et al., 2024). While these frameworks characterize information campaigns and the actors involved, they are limited to the digital realm, and to single actions by single actors. This paper profiles the flow of actions from actors across time, as an information campaign becomes increasingly heated.

Research has increasingly recognized that modern information operations leverage multi-platform digital ecosystems to amplify content reach and strategically target specific audiences (Lukito, 2020). For example, coordinating narratives migrate between social media platforms (Ng et al., 2022), and Russian false information penetrated political subreddits through cross-platform coordination strategies (Hanley et al., 2023). Fewer works studied the crossover of narratives from the online to the offline space, although 2020 study showed the correlation between state mandates and politics and the mask-wearing behavior (Uyheng et al., 2023). This paper looks beyond single temporal moments of campaigns and profiles how narratives mutate as they traverse platform boundaries, fueled by both sets of actors.

Actor analysis on information campaigns mostly focuses on automated actors, i.e. bots, as malicious actors, and profiles their contributions to the spread of false information. Bots make up about 20% of the digital information space, but they generate 30-50% of the posts (Ng & Carley, 2025). Automated campaigns by bots operated by the Russian state during the 2016 US elections revealed systematic manipulation of public discourse (Zannettou et al., 2019), and bots were identified to share problematic information on social media and distort amplification. (Giglietto et al., 2020). Other studies view information campaigns as coordinated operations by automated troll agents operated by the Russian state (Starbird et al., 2019). However, the landscape of actors is not solely limited to bot networks or actors. Recent research reveals a sophisticated ecosystem that includes state-sponsored troll farms and hybrid human-AI operations (Broniatowski et al., 2018; Ng et al., 2024; Uyheng et al., 2022). At the same time, constructive actors like governments, social media platforms and grassroots movements have also banded together to protect the health of the digital ecosystem (Cavelty & Wenger, 2022; Gorwa & Guilbeault, 2020).

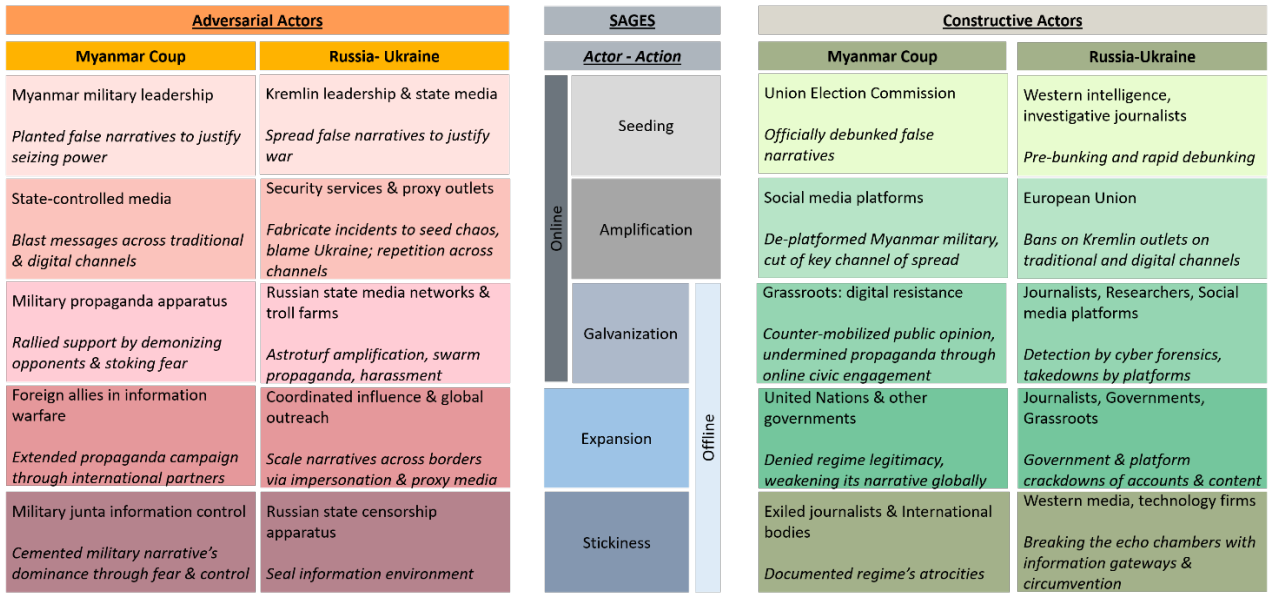

**Figure 1: Actors & Actions of SAGES Framework across two case studies**

This work profiles the involvement of actors in the evolution of a digital narrative into its physical manifestation. Prior frameworks primarily characterize information campaigns in terms of technical tactics (MITRE ATTA&CK, D3FEND), or single actor interventions (BEND) within online environments. These approaches, while valuable, remain limited in their ability to capture the impact of actors on pushing and evolving narratives. Different actor types, from creators, consumers, influencers, endorsers, propagators, and resistors, each propagate narratives differently (George et al., 2022). This paper advances the field by introducing the five-stage **SAGES Framework** that systematically analyzes a narrative's lifecycle, and situates actor interventions within a temporal, cross-platform, and digital-to-physical continuum. Such a framework moves beyond static or platform-bound analyses, offering a lens to understand how influence campaigns unfold and where countermeasures may be most effectively deployed. This work also profiles how distinct adversarial and constructive actors can intervene at critical junctures to accelerate the narrative's progression, redirect its framing, or suppress its spread. This paper expands past work on static actor analysis to capture the dynamic actor participation and influence patterns. Understanding these interventions is essential for diagnosing how online narratives contribute to societal polarization, democratic backsliding, or unrest (Orhan, 2022).

To bridge this gap, this paper builds the SAGES framework, a five-stage model – **Seeding, Amplification, Galvanization, Expansion, and Stickiness.** The SAGES framework traces the trajectory of online narratives from inception to real-world impact. Particularly, this paper studies the evolution of narratives through an **actor-oriented lens**, outlining how both adversarial and constructive actors influence each phase. Next, this paper illustrates the application of this framework through two case studies, situated in Myanmar and Russia-Ukraine, then provides a cross-case analysis. By illuminating these intervention points, the framework provides a lens to analyze narrative manipulation and resilience across digital platforms and sociopolitical contexts.

## 2. SAGES Framework

To understand how influence campaigns evolve across digital ecosystems, this paper applies a five-stage actor-oriented framework—**SAGES**—that tracks narrative progression from introduction to real-world impact. Each stage represents a critical intervention point where different actors—both adversarial and constructive—can shape, accelerate, or disrupt narrative trajectories. Figure 1 visualizes this framework and maps actor interventions across the two case studies.

1. **Seeding** is the initial stage in which a narrative is introduced into the information ecosystem. Adversarial actors may exploit fringe forums, anonymous leaks, or ambiguous events to introduce misleading narratives, while constructive actors may attempt to expose or preemptively debunk such content.

2. **Amplification** marks the rise in visibility and engagement. Here, adversarial actors use bot networks, coordinated influencers, or manipulated media to rapidly spread narratives. Constructive actors counter by deplatforming, fact-checking, and reducing algorithmic virality.

3. **Galvanization** is the turning point where digital discourse shifts toward real-world mobilization. Narratives at this stage begin to provoke strong emotional responses and calls to action. Adversaries often invoke polarizing frames to incite action or outrage, while constructive actors may deploy counter-narratives or pre-bunking interventions.

4. **Expansion** occurs when online mobilization leads to real-world actions—such as protests, unrest, or international propaganda campaigns. Adversarial actors may escalate via global alliances or disinformation partnerships, while constructive actors may respond through diplomatic measures, policy coordination, or civic resistance.

5. **Stickiness** represents the cultural entrenchment of a narrative. Adversaries aim to suppress counter-narratives and normalize their framing over time, while constructive actors seek to preserve informational pluralism and prevent the long-term internalization of falsehoods through media literacy, investigative journalism, and archival fact-checking.

This framework was developed through iterative coding and analysis by three information researchers who have studied and profiled information campaigns within governments and civil society contexts. It builds upon prior work on narrative lifecycle models such as the MITRE framework and the BEND framework (Carley, 2020; Menard et al., 2025), cognitive psychology studies of what stickiness of truths and myths, (Schwarz et al., 2016) and adds a structured, actor-based lens for examining interventions at each stage.

Next, we apply the SAGES framework on two comparative case studies to illustrate how narrative evolution is influenced by actor interventions across different sociopolitical settings.

## 3. Case Study

This section illustrates the actor interventions within the SAGES framework through two case studies: The first is the 2021 Myanmar Coup, where digital disinformation resulted in the erosion of democratic norms. The second is the 2022 Russia invasion of Ukraine, which illustrates how large-scale disinformation campaigns can be weaponized to legitimize interstate aggression. Beyond these case studies, we do a cross-case synthesis to highlight the similarities and differences of the two events. Figure 1 highlights the key actors of the two case studies.

### 3.1. 2021 Myanmar Coup

The 2021 Myanmar military coup illustrates how disinformation campaigns can be engineered to erode societal norms, dismantle democratic institutions, and justify authoritarian power grabs. The SAGES framework traces how the Tatmadaw (Myanmar's military junta) mobilized a full-spectrum influence operation across digital and legacy media. The framework also contrasts these hostile tactics with the fragmented but notable counter-efforts by journalists, activists, and international media organizations (Stokke & Kyaw, 2024).

**Seeding Stage:**

Adversarial Actors. The Seeding stage was characterized by strategic disinformation seeding by the Tatmadaw, Myanmar's military leadership, led by Senior General Min Aung Hlaing. Months before the coup began in February 2021, the military strategically planted false narratives alleging widespread electoral fraud during the November 2020 elections. The Tatmadaw established a "True News" team on Facebook, supposedly to provide accurate news. In reality, the Facebook group was spreading claims of voter fraud. The cornerstone narrative of "8.6 million irregularities in voter lists across 314 townships enabling multiple voting and widespread voting malpractice" laid the groundwork to legitimize their seizure of power (Lidauer, 2023). More so, this narrative seeded deep-rooted distrust against the elected government and justified their military intervention (Tun et al., 2020).

Constructive Actors. To counter these claims, the Myanmar's Union Election Commission (UEC) swiftly issued public rebuttals, clarifying that there was no evidence of election malpractice. This official statement was a strategic narrative engineering to deflate the military's narrative before it could fully take hold, in order to reinforce the legitimacy of the election outcomes (Lidauer, 2023; Strangio, 2021).

**Amplification Stage**

Adversarial Actors. Post-coup, state-controlled media outlets and military-aligned proxies aggressively disseminated the Tatmadaw's narratives. The narratives amplified across controlled and digital channels included accusations against opposition groups, such as labeling all resistance groups as terrorist movements, portraying them as violators of human rights, and claiming that they destroy traditional religious and national values (Mihr, 2023). In response to Facebook bans, the junta quickly migrated the same content to Telegram and TikTok, keeping the propaganda stream alive for hard-line domestic audiences, channeling hate and disinformation (Nachemson, 2021).

Constructive Actors. Social media platforms, particularly Facebook, significantly curbed this amplification by de-platforming military-associated pages and profiles. The profile ban, implemented shortly after the coup, critically reduced the military's ability to coordinate and spread disinformation online (Faxon et al., 2023). Subsequent collaboration with civil-society fact-checkers helped identify and remove dozens of imitation accounts that attempted to evade these restrictions (ASEAN, 2021).

**Galvanization Stage**

Adversarial Actors. During galvanization, the Myanmar military's propaganda apparatus aimed to incite nationalist fervor by labelling opposition forces as "terrorists" and painting protesters as destructive forces threatening national stability. The intent was to provoke public anger and support for military action against pro-democracy groups, therefore legitimizing repression (Mihr, 2023).

Constructive Actors. Grassroots digital resistance emerged as a powerful counterbalance, notably through the #WhatsHappeningInMyanmar online campaign. Activists and ordinary citizens employed digital platforms to document real-time abuses by the military and correct false narratives. This organic movement effectively galvanised global attention and domestic solidarity against the military's misinformation, providing alternative credible narratives to citizens (Phattharathanasut, 2024).

**Expansion Stage**

Adversarial Actors. The junta sought to expand its propaganda globally, notably collaborating with foreign information warfare allies, particularly Russian state media. Such partnerships expanded the military information campaign through the formation and leverage of tactical and international alliances. These actions aimed at enhancing the perceived legitimacy of the coup internationally and spreading pro-junta narratives in the global information ecosystem (Golovchenko et al., 2018).

Constructive Actors. The international community responded to the junta's actions by diplomatically isolating the junta. The United Nations maintained Myanmar's existing ambassador who openly defied the military, rejecting the junta's legitimacy at the international level. This diplomatic strategy effectively undermined the junta's narrative abroad, reinforcing global awareness and condemnation of the coup (Passeri, 2025).

**Stickiness Stage**

Adversarial Actors. To ensure lasting influence, the junta implemented extreme censorship measures to suppress counter-speech. The junta revoked licenses of independent media, arrested journalists, and established stringent internet controls. This suppression sought to embed their narrative in the national consciousness by eradicating alternative voices, therefore securing the stickiness of their ideal narrative (Padmanabhan et al., 2021).

Constructive Actors. Exiled independent media and international organizations played crucial roles in sustaining factual counter-narratives. Media outlets like Myanmar-Now and The Irrawaddy established systematic fact-checking efforts, providing continuous exposure and archiving of military abuses and falsehoods. These persistent truth-telling campaigns have significantly disrupted the junta's efforts to achieve long-term narrative dominance, ensuring that alternative and truthful accounts remain accessible to global and local audiences (Thu, 2024).

### 3.2. 2022 Russia Invasion of Ukraine

Russia's 2022 invasion of Ukraine illustrates the strategic use of information warfare to provide narrative justification for territorial aggression. aggression. This case study examines how the Kremlin systematically deployed narrative manipulation to manufacture pretexts for the invasion, discredit Ukrainian sovereignty, and rally domestic and international support. It also highlights the countervailing efforts of Ukrainian civil society, independent media, and global fact-checking coalitions in exposing, countering, and mitigating these adversarial narratives.

**Seeding Stage**

Adversarial Actors. In the months preceding Russia's invasion of Ukraine in February 2022, Russian state leadership—primarily President Vladimir Putin and senior officials—carefully crafted and seeded narratives that framed Ukraine as being controlled by neo-Nazis who were committing genocide against Russian-speaking populations in the Donbas region. Putin explicitly invoked terms such as "genocide" to rationalize military aggression, while Russia's Defence Ministry amplified conspiracies alleging planned Ukrainian and NATO provocations (Rossoliński-Liebe & and Willems, 2022).

Constructive Actors. Pre-emptive countermeasures, known as "pre-bunking," significantly disrupted the adversarial narratives. Western intelligence agencies, particularly those from the US and UK, publicly released declassified intelligence exposing Russia's planned false-flag operations and disinformation strategies (Pelley, 2022). These disclosures weakened the narrative foundation for war by forewarning global audiences and governments about Russia's manipulative intent (Phythian & Strachan-Morris, 2024).

**Amplification Stage**

Adversarial Actors. Post-invasion, Russian state media outlets—most notably RT, Sputnik, and state-controlled television channels—engaged in intensive amplification of Kremlin narratives. A coordinated "firehose of falsehood" strategy was employed, saturating global digital and broadcast channels with false or misleading content (Paul & Matthews, 2016). This strategy aimed at legitimizing the invasion, demonizing Ukraine, and shifting blame onto NATO and the West for regional instability.

Complementing official channels, Russia mobilized extensive troll farms and fake digital personas, notably via organizations like the Internet Research Agency (Linvill et al., 2019; Ng & Taeihagh, 2021). These entities utilized bot networks, manipulated social media algorithms, and even set up deceptive websites impersonating credible news sources (Akhtar et al., 2024). Such tactics dramatically amplified pro-Kremlin content across various digital platforms (Marigliano et al., 2024).

Constructive Actors. Rapid counter-responses curtailed the adversarial efforts effectively. The European Union decisively banned RT and Sputnik broadcasts within its territory, significantly reducing the immediate impact and reach of Russian narratives (Chee & Chee, 2022). Simultaneously, major tech platforms including X, Facebook, and Google took aggressive actions—such as down-ranking, labelling, and removing Russian state-linked disinformation—to diminish its prominence and credibility in online spaces (Culliford, 2022).

**Galvanization Stage**

Adversarial Actors. At the Galvanizations stage, pro-Kremlin digital activists and nationalist influencers orchestrated large-scale online campaigns to reinforce domestic support for the invasion. Notably, groups such as Cyber Front Z mobilized volunteers and paid trolls, flooding online spaces with coordinated pro-war content and employing harassment tactics against opponents. These campaigns were designed to create a facade of grassroots support, thereby bolstering nationalistic fervor within Russia and intimidating dissenters abroad (Alieva et al., 2024; Marigliano et al., 2024).

Constructive Actors. Effective countermeasures included platform-level crackdowns and grassroots digital resistance. Western governments shared intelligence about Russian troll operations with tech platforms, leading to widespread takedowns of coordinated accounts (Alieva et al., 2024; Marigliano et al., 2024). Additionally, the spontaneous emergence of digital activist collectives, exemplified by the North Atlantic Fella Organization (NAFO), successfully disrupted and mocked pro-Kremlin narratives. NAFO's meme-driven campaigns undermined Russian digital propaganda by consistently exposing and ridiculing its absurdities, helping to reduce its psychological impact (Welle, 2022).

**Expansion Stage**

Adversarial Actors. Russia systematically expanded its digital propaganda operations beyond traditional Western audiences, targeting regions in the Global South, particularly Africa and the Middle East (Bond, 2022). Russian Foreign Minister Sergey Lavrov and state media outlets exploited pre-existing anti-colonial sentiments, framing Western sanctions—not Russia's aggression—as the primary cause of global food and energy crises (Pathé Duarte, 2024). This narrative found resonance in regions with historical skepticism towards Western motives, thus extending Russia's influence globally (Janadze, 2022).

Constructive Actors. Western diplomatic and strategic communication initiatives actively countered Russia's international outreach (Bond, 2022). The US State Department and European-Union diplomatic corps engaged directly with African and Middle Eastern countries, systematically debunking Russian claims and clarifying the true sources of global crises—particularly Russia's naval blockade of Ukrainian grain exports. Additionally, efforts were made to bolster local independent media capacity in target regions, ensuring accurate information reached vulnerable demographic audiences (Janadze, 2022).

**Stickiness Stage**

Adversarial Actors. To ensure the permanence of their propaganda, Russian authorities employed extreme internal censorship measures. New legislation criminalized reporting contradictory to official narratives, with severe penalties designed to suppress independent journalism (Sherstoboeva, 2024). State-controlled media relentlessly repeated Kremlin messaging, aiming to cement disinformation into the collective consciousness of Russian citizens by eliminating alternative information sources (Cain, 2022).

Constructive Actors. Creative circumvention strategies and persistent external informational support significantly challenged these efforts. Western and independent Russian journalists utilized innovative communication channels—including shortwave radio, Virtual Private Networks (VPNs), encrypted messaging apps like Telegram, and diaspora-focused platforms—to maintain information flows into Russia. Tech companies like Meta continued applying fact-check labels and refused to comply with censorship demands, albeit at the cost of restrictions within Russia (Bond, 2022). These ongoing countermeasures preserved an avenue of truthful reporting for Russians determined to seek it, thereby limiting the ultimate "stickiness" of Kremlin disinformation (Culliford, 2022).

### 3.3. Cross-Case Synthesis

This section synthesizes insights from the information campaigns of the 2021 Myanmar Coup and 2022 Russian invasion of Ukraine that were profiled using the SAGES framework. The similarities and differences between the adversarial tactics and counter-influence is highlighted in Figure 1. A narrative synthesis of the figure follows:

In terms of **Seeding**, both cases began with meticulously crafted false narratives aimed at justifying actions that were widely viewed as illegitimate internationally. Myanmar's military propagated claims of electoral fraud to rationalize the coup, whereas Russia disseminated allegations of neo-Nazi control and genocide in Ukraine to justify the invasion. In both scenarios, state authorities were the primary instigators, strategically leveraging pre-existing societal fears and grievances to enhance the acceptance of their designed narrative within the domestic context (Lidauer, 2023).

At the **Amplification** stage, both campaigns employed extensive digital and traditional media resources, harnessing state-controlled media and coordinated digital networks to saturate the information environment. Myanmar relied significantly on state-run media broadcasters and proxies after restrictions were imposed by global social media platforms (Kim & Kim, 2025). Russia demonstrated greater digital sophistication, operating complex troll farms and deploying large-scale astroturfing to reinforce its global narratives through fake news sites and manipulative social media tactics (Marigliano et al., 2024). While both cases are similar in tactics, the scale and sophistication of Russia's digital amplification efforts exceeded Myanmar's.

The cases diverged slightly in the execution of **Galvanization**, but shared the core tactic of mobilizing nationalist sentiment against portrayed enemies. Myanmar's military explicitly labelled opposition as "terrorists," galvanizing domestic audiences towards active or passive support of military suppression. Russia similarly invoked nationalistic fervor via pro-war digital brigades such as Cyber Front Z, aiming to stimulate public backing for the invasion and marginalize dissenting voices through coordinated online harassment. Interestingly, grassroots and spontaneous digital resistance, such as Myanmar's online #WhatsHappeningInMyanmar campaign and Ukraine-related NAFO activism, emerged as effective counters in both cases (Phattharathanasut, 2024). These movements effectively diluted the adversarial narratives by directly engaging and challenging them through credible, citizen-driven information.

During the **Expansion** stage, both states expanded their propaganda internationally but with varying methodologies. Myanmar sought direct cooperation with international authoritarian partners, notably Russian state media, to export and legitimize its narrative externally (Irrawaddy, 2023). Russia, leveraging its far-reaching diplomatic and media networks, specifically targeted Global South countries, exploiting historical anti-colonial sentiments to shift blame for global crises onto Western nations (DFRL, 2024). In response, international isolation tactics against Myanmar and strategic diplomatic outreach and localized media empowerment against Russian narratives proved successful countermeasures, significantly limiting narrative acceptance beyond immediate spheres of influence (Irrawaddy, 2023).

Finally, the **Stickiness** stage revealed starkly parallel approaches, with both regimes establishing stringent domestic censorship to entrench disinformation narratives. Myanmar's junta revoked licenses, criminalized independent journalism, and censored internet usage, effectively attempting to erase dissenting narratives (Freedom House, 2024). Russia, similarly, imposed harsh legal penalties against any deviation from official war narratives and restricted independent and international media access domestically (Iashchenko, 2023). Constructive actors across both cases adopted innovative circumvention strategies, including the usage of VPNs, shortwave radios, and encrypted platforms like Telegram, complemented by consistent fact-checking and investigative journalism. These persistent efforts limited long-term narrative entrenchment, preserving access to truthful information for resilient audience segments.

## 5. Implications for Policy, Technology, and Research Communities

Understanding the actor-driven dynamics of narrative evolution offers a roadmap for identifying where strategic interventions are most likely to disrupt adversarial influence campaigns. Each stage in the SAGES framework presents a distinct set of vulnerabilities that can be exploited or countered by policymakers, platform designers, and researchers.

In the **Seeding stage**, when a narrative is first introduced, intervention is most effective when focused on origin detection and exposure. Policymakers should support the development of early-warning systems that monitor fringe forums and online spaces where narratives are likely to be seeded via strategic leaks. Technology platforms can implement anomaly detection tools that flag synthetic account activity or high-risk clusters of coordinated content. Academic researchers can contribute by profiling common seeding signatures and identifying early indicators of coordinated narrative injection.

During the **Amplification stage**, where narratives gain traction through both organic humans and inorganic bots, interventions should focus on disruption of virality and surfacing of context. Platform architects can deploy algorithmic friction such as repost delays, warning labels, or rate-limiting mechanisms for unverified high-velocity content. Governments can mandate disclosure of coordinated promotional campaigns and ensure public transparency in platform moderation practices. Researchers can support these interventions by studying platform-specific affordances and designing real-time detection tools that distinguish organic from inauthentic amplification.

The **Galvanization stage** marks a shift toward mobilization. At this point, interventions must target coordination pathways and emotional intensity. Fact-checkers, supported by platform integrations, can launch pre-bunking and rapid response debunking to counter emotionally resonant falsehoods. Technology providers should monitor spikes in mobilization language or calls to action across platforms, especially when originating from suspected inauthentic networks. Academic teams can refine predictive models that map narrative evolution from emotional resonance to real-world planning behaviors.

Intervening in the **Expansion stage**, where digital narratives inspire real-world action, requires integrated policy and technical coordination. Governments should establish digital crisis response cells that combine law enforcement, cyber agencies, and media liaison teams to respond to online-triggered unrest. Technology platforms can collaborate by sharing anonymized engagement trend data with designated crisis partners under clear governance protocols. Academic researchers can enhance models that forecast escalation risks based on multi-platform narrative tracking.

In the **Stickiness stage**, where narratives become culturally embedded or cyclically resurface, the emphasis should shift to long-term resilience and narrative decay. Policymakers should integrate critical thinking and media literacy into national education systems and fund community-based awareness campaigns. Platforms can support archival fact-checking initiatives and use historical correction labels to reduce the recirculation of debunked narratives. Academics can conduct longitudinal studies on the lifecycle of persistent falsehoods and identify the psychological or cultural conditions that sustain them.

While many interventions are stage-specific, several institutional actors play a cross-cutting role across the entire narrative lifecycle. Fact-checking organizations disrupt both early virality and long-term stickiness through scalable verification pipelines, searchable public archives, and educational outreach. Inter-governmental task forces enhance response capabilities by enabling intelligence sharing, coordinated attribution of foreign-origin campaigns, and harmonized counter-messaging across jurisdictions. Civil society organizations and public broadcasters reinforce democratic resilience by delivering credible counter-narratives, fostering public trust, and maintaining visibility of vetted information in moments of crisis or ambiguity.

It is important to acknowledge that these interventions may carry trade-offs. Mechanisms such as algorithmic friction, content labeling, or coordinated counter-messaging must be carefully balanced with safeguards for freedom of expression, transparency, and the risk of overreach or false positives (Gorwa et al., 2020). Effective narrative governance depends not only on disrupting manipulation but also on preserving the openness and pluralism of digital discourse (Ng & Carley, 2025).

## 6. Conclusions

This paper presented an actor-oriented perspective of the evolution of narratives from the digital to physical space with the SAGES framework. This offers a structured lens to analyze how adversarial and constructive actors shape the evolution of narratives across digital and physical information spaces. By tracing narrative trajectories through the stages of Seeding, Amplification, Galvanization, Expansion, and Stickiness, we highlight the strategic roles actors play in manipulating, redirecting, or countering narratives. Through comparative case studies—the 2021 Myanmar coup and the 2022 Russia–Ukraine war—we demonstrate the versatility of the SAGES framework in dissecting narrative influence operations across different geopolitical contexts.

Our findings underscore that influence campaigns are not driven solely by automated amplification or singular actor types, but by dynamic and coordinated interventions across actor ecosystems. Recognizing these actor-specific intervention points equips policymakers, platform designers, and operational analysts with a clearer roadmap to anticipate and disrupt disinformation trajectories before they become embedded in public consciousness.

**Future Directions.** Building on the stage-specific research gaps outlined earlier, this framework opens several avenues for further investigation. These include developing metrics to quantify actor influence at each SAGES stage, simulating multi-actor interactions in synthetic media environments, and testing the framework's generalizability across culturally diverse information ecosystems. Emerging challenges—such as the rise of AI-generated content and synthetic personas—warrant particular attention, especially in contexts with low digital literacy or weak information infrastructure.

## 7. References

Akhtar, M. M., Masood, R., Ikram, M., & Kanhere, S. S. (2024). SoK: False Information, Bots and Malicious Campaigns: Demystifying Elements of Social Media Manipulations. *Proceedings of the 19th ACM Asia Conference on Computer and Communications Security*, 1784–1800. https://doi.org/10.1145/3634737.3644998

Alieva, I., Kloo, I., & Carley, K. M. (2024). Analyzing Russia's propaganda tactics on Twitter using mixed methods network analysis and natural language processing: A case study of the 2022 invasion of Ukraine. *EPJ Data Science*, *13*(1), Article 1. https://doi.org/10.1140/epjds/s13688-024-00479-w

Bond, S. (2022, September 27). Facebook takes down Russian network impersonating European news outlets. *NPR*. https://www.npr.org/2022/09/27/1125217316/facebook-takes-down-russian-network-impersonating-european-news-outlets

Cain, S. (2022, March 4). BBC website blocked in Russia as shortwave radio brought back to cover Ukraine war. *The Guardian*. https://www.theguardian.com/media/2022/mar/04/bbc-website-blocked-in-russia-as-shortwave-radio-brought-back-to-cover-ukraine-war

Chee, F. Y., & Chee, F. Y. (2022, March 2). EU bans RT, Sputnik over Ukraine disinformation. *Reuters*. https://www.reuters.com/world/europe/eu-bans-rt-sputnik-banned-over-ukraine-disinformation-2022-03-02/

Culliford, E. (2022, February 28). *Twitter will label, reduce visibility of tweets linking to Russian state media | Reuters*. https://www.reuters.com/technology/twitter-will-label-reduce-visibility-tweets-linking-russian-state-media-2022-02-28/

DFRL. (2024, February 29). Undermining Ukraine: How Russia widened its global information war in 2023. *Atlantic Council*. https://www.atlanticcouncil.org/in-depth-research-reports/report/undermining-ukraine-how-russia-widened-its-global-information-war-in-2023/

Faxon, H. O., Kintzi, K., Tran, V., Wine, K. Z., & Htut, S. Y. (2023). Organic online politics: Farmers, Facebook, and Myanmar's military coup. *Big Data & Society*, *10*(1), 20539517231168101. https://doi.org/10.1177/20539517231168101

Freedom House. (2024). *Myanmar: Freedom on the Net 2024 Country Report*. Freedom House. https://freedomhouse.org/country/myanmar/freedom-net/2024

Giglietto, F., Righetti, N., Rossi, L., & Marino, G. (2020). It takes a village to manipulate the media: Coordinated link sharing behavior during 2018 and 2019 Italian elections. *Information, Communication & Society*, *23*(6), 867–891. https://doi.org/10.1080/1369118X.2020.1739732

George, J., Gerhart, N., & Torres, R. (2023). Uncovering the truth about fake news: A research model grounded in multi-disciplinary literature. *Fake News on the Internet*, 175-202.

Golovchenko, Y., Hartmann, M., & Adler-Nissen, R. (2018). State, media and civil society in the information warfare over Ukraine: Citizen curators of digital disinformation. *International Affairs*, *94*(5), 975–994. https://doi.org/10.1093/ia/iiy148

Gorwa, R., Binns, R., & Katzenbach, C. (2020). Algorithmic content moderation: Technical and political challenges in the automation of platform governance. *Big Data & Society*, *7*(1), 2053951719897945. https://doi.org/10.1177/2053951719897945

Hanley, H. W. A., Kumar, D., & Durumeric, Z. (2023). Happenstance: Utilizing Semantic Search to Track Russian State Media Narratives about the Russo-Ukrainian War on Reddit. *Proceedings of the International AAAI Conference on Web and Social Media*, *17*, 327–338. https://doi.org/10.1609/icwsm.v17i1.22149

Iashchenko. (2023, September 18). Russian disinformation about the Ukrainian conflict since 2014: Fact-checking and recurring patterns. *Aspenia Online*. https://aspeniaonline.it/russian-disinformation-about-the-ukrainian-conflict-since-2014-fact-checking-and-recurring-patterns/

Irrawaddy, T. (2023, September 7). Myanmar Junta Adds Russia's 'Firehose of Propaganda' to its Arsenal. *The Irrawaddy*. https://www.irrawaddy.com/news/myanmars-crisis-the-world/myanmar-junta-adds-russias-firehose-of-propaganda-to-its-arsenal.html

Janadze, E. (2022). *Russian cyber strategy in the Middle East and North Africa (MENA): Analysing the Kremlin's disinformation efforts amid 2022 invasion of Ukraine*. https://dspace.cuni.cz/handle/20.500.11956/178359

Kim, D. K. D., & Kim, I. (2025). Social Media as a Seed of Connective Democracy in Myanmar (Burma): Freedom of Speech, Contractarianism, and Strategic Use of Social Media. *Social Media + Society*, *11*(1), 20563051251329996. https://doi.org/10.1177/20563051251329996

Kruijver, K., Finlayson, N. B., Cadet, B., & van de Meer, S.(2025). The disinformation lifecycle: an integrated understanding of its creation, spread and effects. *Discover Global Society*, 3(1), 58.

Lidauer, M. (2023). Myanmar's Menu of Electoral Manipulation: Self- and External Legitimation after the 2021 Coup. *Critical Asian Studies*, *55*(3), 397–423. https://doi.org/10.1080/14672715.2023.2212366

Linvill, D. L., Boatwright, B. C., Grant, W. J., & Warren, P. L. (2019). "THE RUSSIANS ARE HACKING MY BRAIN!" investigating Russia's internet research agency twitter tactics during the 2016 United States presidential campaign. *Computers in Human Behavior*, *99*, 292–300. https://doi.org/10.1016/j.chb.2019.05.027

Lukito, J. (2020). Coordinating a Multi-Platform Disinformation Campaign: Internet Research Agency Activity on Three U.S. Social Media Platforms, 2015 to 2017. *Political Communication*, *37*(2), 238–255. https://doi.org/10.1080/10584609.2019.1661889

Marigliano, R., Ng, L. H. X., & Carley, K. M. (2024). Analyzing digital propaganda and conflict rhetoric: A study on Russia's bot-driven campaigns and counter-narratives during the Ukraine crisis. *Social Network Analysis and Mining*, *14*(1), 170. https://doi.org/10.1007/s13278-024-01322-w

Menard, P., Reyes, E., & Bateman, R. (2025). *Understanding Zero Trust Security Implementations via the MITRE ATT&CK and D3FEND Frameworks: Uncovering Trends Across a Decade of Breaches*. https://hdl.handle.net/10125/109071

Mihr, A. (2023). Empires of Eurasia: How imperial legacies shape international security; The neighborhood effect: the imperial roots of regional fracture in Eurasia. *International Affairs*, *99*(6), 2538–2540. https://doi.org/10.1093/ia/iiad274

Ng, L. H. X., & Carley, K. M. (2025). A global comparison of social media bot and human characteristics. *Scientific Reports*, *15*(1), 10973. https://doi.org/10.1038/s41598-025-96372-1

Ng, L. H. X., Cruickshank, I. J., & Carley, K. M. (2022). Coordinating Narratives Framework for cross-platform analysis in the 2021 US Capitol riots. *Computational and Mathematical Organization Theory*. https://doi.org/10.1007/s10588-022-09371-2

Ng, L. H. X., & Taeihagh, A. (2021). How does fake news spread? Understanding pathways of disinformation spread through APIs. *Policy & Internet*, poi3.268. https://doi.org/10.1002/poi3.268

Orhan, Y. E. (2022). The relationship between affective polarization and democratic backsliding: Comparative evidence. *Democratization*, *29*(4), 714–735. https://doi.org/10.1080/13510347.2021.2008912

Padmanabhan, R., Filastò, A., Xynou, M., Raman, R. S., Middleton, K., Zhang, M., Madory, D., Roberts, M., & Dainotti, A. (2021). A multi-perspective view of Internet censorship in Myanmar. *Proceedings of the ACM SIGCOMM 2021 Workshop on Free and Open Communications on the Internet*, 27–36. https://doi.org/10.1145/3473604.3474562

Passeri, A. (2025). Myanmar's Post-Coup Foreign Policy and Alignment Behavior: Assessing the Agency of a "Pariah State." *Journal of Current Southeast Asian Affairs*, *44*(1), 102–124. https://doi.org/10.1177/18681034241303170

Pathé Duarte, F. (2024). Information Disorder and Civil Unrest Russian Weaponization of Social Media Platforms in Mali and Burkina Faso – 2020–2022. *African Security*, *17*(3–4), 205–223. https://doi.org/10.1080/19392206.2024.2423139

Paul, C., & Matthews, M. (2016). *The Russian "Firehose of Falsehood" Propaganda Model: Why It Might Work and Options to Counter It*. https://www.rand.org/pubs/perspectives/PE198.html

Pelley, S. (2022, August 21). *Bellingcat: The online investigators tracking alleged Russian war crimes in Ukraine - 60 Minutes - CBS News*. https://www.cbsnews.com/news/bellingcat-russia-putin-ukraine-60-minutes-2022-08-21/

Phattharathanasut, T. (2024). #WhatsHappeningInMyanmar: The Evolution of the Digital Fight Against Authoritarian State Repression. *International Journal of Communication*, *18*(0), Article 0.

Phythian, M., & and Strachan-Morris, D. (2024). Intelligence & the Russo-Ukrainian war: Introduction to the special issue. *Intelligence and National Security*, *39*(3), 377–385. https://doi.org/10.1080/02684527.2024.2330132

Rossoliński-Liebe, G., & and Willems, B. (2022). Putin's Abuse of History: Ukrainian 'Nazis', 'Genocide', and a Fake Threat Scenario. *The Journal of Slavic Military Studies*, *35*(1), 1–10. https://doi.org/10.1080/13518046.2022.2058179

Sherstoboeva, E. (2024). Russian Bans on 'Fake News' about the war in Ukraine: Conditional truth and unconditional loyalty. *International Communication Gazette*, *86*(1), 36–54. https://doi.org/10.1177/17480485231220141

Schwarz, N., Newman, E., & Leach, W. (2016). Making the truth stick & the myths fade: Lessons from cognitive psychology. *Behavioral Science & Policy*, 2(1), 85-95

Starbird, K., Arif, A., & Wilson, T. (2019). Disinformation as Collaborative Work: Surfacing the Participatory Nature of Strategic Information Operations. *Proceedings of the ACM on Human-Computer Interaction*, *3*(CSCW), 1–26. https://doi.org/10.1145/3359229

Stokke, K., & Kyaw, N. N. (2024). Revolutionary resistance against full autocratization. Actors and strategies of resistance after the 2021 military coup in Myanmar. *Political Geography*, *108*, 103011. https://doi.org/10.1016/j.polgeo.2023.103011

Strangio, S. (2021, January 29). *Amid Coup Fears, Myanmar's Election Commission Rejects Army Election Fraud Claims*. The Diplomat. https://thediplomat.com/2021/01/amid-coup-fears-myanmars-election-commission-rejects-army-election-fraud-claims/

Thu, T. A. (2024, July 16). *Under the junta's shadow: The rise of citizen journalism in Myanmar | Reuters Institute for the Study of Journalism*. https://reutersinstitute.politics.ox.ac.uk/under-juntas-shadow-rise-citizen-journalism-myanmar

Tun, T., Thet, S. A., & Chann, N. (2020, June 10). *Critics Warn of Deception as Myanmar Military Returns to Facebook After 2018 Purge*. Radio Free Asia. https://www.rfa.org/english/news/myanmar/military-facebook-06102020170558.html

Welle, D. (2022, September 19). *#NAFO: Ukraine's info warriors battling Russian trolls | ABS-CBN*. https://www.abs-cbn.com/spotlight/09/19/22/nafo-ukraines-info-warriors-battling-russian-trolls

Zannettou, S., Caulfield, T., De Cristofaro, E., Sirivianos, M., Stringhini, G., & Blackburn, J. (2019). Disinformation Warfare: Understanding State-Sponsored Trolls on Twitter and Their Influence on the Web. *Companion Proceedings of The 2019 World Wide Web Conference on - WWW '19*, 218–226. https://doi.org/10.1145/3308560.3316495